\documentclass[RNAAS]{aastex62}

\begin{document}

\title{Updated proper motions for Local Group dwarf galaxies using
  Gaia Early Data Release 3}


\author{Alan W. McConnachie}
\correspondingauthor{Alan W. McConnachie}
\email{alan.mcconnachie@nrc-cnrc.gc.ca}
\affil{NRC Herzberg Astronomy and Astrophysics, 5071 West Saanich
  Road, Victoria, B.C., Canada, V9E 2E7}
\author{Kim A. Venn} 
\affil{Physics \& Astronomy Department, University of Victoria, 3800 Finnerty Rd, Victoria, B.C., Canada, V8P 5C2}

\begin{abstract}
  Updated systemic proper motion estimates for 58 Milky Way satellite galaxies, based on Gaia Early Data
  Release 3 (EDR3), are
  provided. This sample is identical to that studied by
  \cite{mcconnachie2020a} and the methodology is essentially unchanged from the
  original paper. The superiority of Gaia EDR3 compared to Gaia Data
  Release 2
  means that Bo{\"o}tes 4, Cetus 3, Pegasus 3 and Virgo 1 have detectable
  systemic proper motions for the first time. For the entire galaxy
  sample, the median random uncertainties in the systemic proper
  motions are approximately a factor of two better than the previous estimates using Gaia
  DR2. Relevant systematic errors, which are also a factor of two
  smaller, dominate over random uncertainties for
  25 out of the 58 objects in the sample.
\end{abstract}

\keywords{
Galaxy: halo --- 
galaxies: dwarf --- galaxies: general --- Local Group}

\section{}

Gaia Data Release 2 (DR2; \citealt{gaia2018b}) revolutionized our
ability to measure the systemic proper motions of relatively distant Milky Way
satellites (e.g., \citealt{helmi2018b, massari2018, simon2018, simon2020, fritz2018,
  fritz2019,kallivayalil2018, pace2019, longeard2018, longeard2020a, longeard2020b,
  mau2020, mcconnachie2020a}). Gaia Early Data Release 3
(EDR3: \citealt{brown2020}) now provides stellar proper motion measurements that are typically a factor of 2
more precise than for DR2. Here, we provide
updated values for the systemic proper motions of most of the Milky Way dwarf galaxies (including candidate systems)
using Gaia EDR3 and the technique of
\citet[hereafter MV2020]{mcconnachie2020a}.

We refer to MV2020 for information
regarding the methodology used to derive systemic proper motion
estimates. Almost all of the current analysis is identical to the original
paper, and this includes using the same galaxy parameters and ancillary radial velocity
data as tabulated in MV2020. The only difference is
that we use Gaia EDR3 instead of DR2. As such, we make the following
modifications relative to MV2020:

\begin{itemize}
\item we correct stellar parallaxes for the global mean offset of
  $-0.017$\,mas, as determined in \cite{lindegren2020b};
\item we select only those stars with a
  corrected flux excess factor, $|C^*| \le 3 \sigma_{C^*}(G)$, following
  the definitions and notation of \cite{riello2020} (their Equations 6 and 18);
\item we continue using {\tt ruwe}$<1.3$ (see, e.g., \citealt{lindegren2020a}).
\end{itemize}

Table 1 is an update of
Table 4 from MV2020 and provides the names and derived proper motions of
all the galaxies considered in this contribution. Tangential and
radial Galactocentric velocities are also given, using the conversions detailed in
MV2020. Error bars refer to random uncertainities only and are 16th/84th percentiles. \cite{lindegren2020a} estimate an RMS value of
33\,$\mu$as\,yr$^{-1}$ per component for the systematic uncertainity
in the proper motions on spatial scales of $\le 0.125^\circ$. This is
broadly the same spatial scale as many of the Milky Way dwarf galaxies. For DR2,
this value was 66\,$\mu$as\,yr$^{-1}$. We note that, for the new measurements, the typical
  systematic uncertainty is larger than the
  random uncertainty in at least one direction for 25 out of the 58
  objects considered. 

We make the following brief comments on the values presented in Table 1:

\begin{itemize}
\item We present first estimates of the systemic proper motions of
  Bo{\"o}tes 4, Cetus 3, Pegasus 3 and Virgo 1, although only a single
  star is identified as a member ($P_{sat} \ge 0.5$) in each of the latter three. The proper motion of Indus 2 remains
  undetectable in Gaia EDR3 using this technique.
\item The MV2020 estimates for Cetus 2, Pictor 2 and Leo T were based on DR2 data
  with looser quality cuts compared to other systems. The
  new estimates for these objects use the same quality cuts as the rest
  of the current sample.
\item The median random uncertainity in the systemic proper motions
  for the whole sample is 0.05\,mas\,yr$^{-1}$, compared to 0.08\,mas\,yr$^{-1}$ for
  MV2020. Six systems (Antlia 2, Cetus 2, Indus
  1, Leo 1, Pictor 2 and Segue 1) have a factor
  of three or better improvement in their random uncertainties. Cetus 2
  has the most dramatic improvement of an order of magnitude. 
\item The new estimates are generally within a few
  standard deviations of the previous DR2 estimates, especially
  when systematic uncertainties are considered. The most significantly
  different estimates are for Segue 1, Segue 2, and
  Triangulum 2. Examination of the new estimates for these
  systems suggest they are robust, and in all cases more stars are
  associated as members ($P_{sat} \ge 0.5$ ) than before. Indeed, all targets in Segue 1
  and Triangulum 2 that also have high resolution spectroscopy (see
  Tables 6 and A1 in MV2020) now have membership
  probabilities of $P_{sat} \ge 0.5$. 
\item Previously, 75 out of 80 stars in these systems
  with high resolution spectroscopy were assigned as
  members by MV2020 (their Tables 6 and A1) if they were located at $r \le
  4 r_h$ and were not horizontal branch stars. Using the new estimates, all 80 stars
  are assigned as members. In fact, all 83 stars within $5.2\,r_h$
  that are not on the horizontal branch
  are assigned as members.
\item For some of the faintest and most distant systems (Bo{\"o}tes 4,
  Cetus 3, Eridanus 2,
  Eridanus 3, Leo T, Pegasus 3, Pisces 2, and Reticulum 3), the formal uncertainties on
  the proper motions are affected significantly by the width of the Gaussian prior. This prior favors estimates in which the
  galaxies are bound to the Milky Way. For the three most distant galaxies - Eridanus
  2, Leo T and Phoenix - this prior is questionable. As
  such, we also present the systemic proper motion estimates of these
  three galaxies using
  a uniform prior in the last three lines of Table 1.
\end{itemize}

The systemic proper motions in Table 1 will be incorporated into
future updates of the \cite{mcconnachie2012} nearby galaxy catalog,
available
online\footnote{\url{http://www.astro.uvic.ca/~alan/Nearby_Dwarf_Database.html}}. Also
included at this URL will be the updated systemic proper motion
estimates using Gaia EDR3 for some of the more distant Local Group
dwarf galaxies (IC1613, Leo A, NGC 6822 and WLM), as recently
presented in \cite{mcconnachie2020b}.
 
\begin{table*} {\scriptsize
\begin{center}
\begin{tabular*}{0.9\textwidth}{l|rrrrrr}
Galaxy &$\mu_\alpha\cos\delta$ (mas\,yr$^{-1}$)&$\mu_\delta$ (mas\,yr$^{-1}$)& $v_\alpha\cos\delta$ (km\,s$^{-1}$) & $v_\delta$ (km\,s$^{-1}$)& $v_t$ (km\,s$^{-1}$)& $v_r$ (km\,s$^{-1}$)\\
\hline 
Antlia2 & $ -0.09 \pm 0.01 $ & $ 0.12 \pm 0.01 $ & $ -27 \pm 6 $ &  $ 122 \pm 6 $ & 125 & 55 \\ 
Aquarius2 & $ -0.17 \pm 0.1 $ & $ -0.43 \pm 0.08 $ & $ -156 \pm 51 $ &  $ -20 \pm 41 $ & 157 & 46 \\ 
Bo{\"o}tes1 & $ -0.39 \pm 0.01 $ & $ -1.06 \pm 0.01 $ & $ 36 \pm 3 $ &  $ -152 \pm 3 $ & 156 & 107 \\ 
Bo{\"o}tes2 & $ -2.33^{+0.09}_{-0.08} $ & $ -0.41 \pm 0.06 $ & $ -302^{+18}_{-16} $ &  $ 101 \pm 12 $ & 319 & -116 \\ 
Bo{\"o}tes4 & $ -0.15 \pm 0.1 $ & $ -0.1 \pm 0.1 $ & $ 12 \pm 99 $ &  $ 10 \pm 99 $ & 15 & --- \\ 
CanesVenatici1 & $ -0.11 \pm 0.02 $ & $ -0.12 \pm 0.02 $ & $ 38 \pm 21 $ &  $ 58 \pm 21 $ & 69 & 80 \\ 
CanesVenatici2 & $ -0.15 \pm 0.07 $ & $ -0.27 \pm 0.06 $ & $ 29 \pm 53 $ &  $ -13 \pm 45 $ & 31 & -94 \\ 
Carina & $ 0.53 \pm 0.01 $ & $ 0.12 \pm 0.01 $ & $ 175 \pm 5 $ &  $ 66 \pm 5 $ & 187 & -2 \\ 
Carina2 & $ 1.88 \pm 0.01 $ & $ 0.13 \pm 0.02 $ & $ 268 \pm 2 $ &  $ -14 \pm 3 $ & 268 & 245 \\ 
Carina3 & $ 3.12 \pm 0.05 $ & $ 1.54^{+0.06}_{-0.07} $ & $ 358 \pm 7 $ &  $ 167^{+8}_{-9} $ & 395 & 52 \\ 
Centaurus1 & $ -0.14 \pm 0.05 $ & $ -0.2 \pm 0.04 $ & $ 58 \pm 28 $ &  $ -35 \pm 22 $ & 68 & --- \\ 
Cetus2 & $ 2.84^{+0.05}_{-0.07} $ & $ 0.46^{+0.06}_{-0.07} $ & $ 205^{+6}_{-9} $ &  $ 247^{+7}_{-9} $ & 321 & --- \\ 
Cetus3 & $ 0.14 \pm 0.08 $ & $ -0.16 \pm 0.08 $ & $ 7 \pm 95 $ &  $ -10 \pm 95 $ & 13 & --- \\ 
Columba1 & $ 0.19 \pm 0.06 $ & $ -0.36 \pm 0.06 $ & $ 38 \pm 52 $ &  $ -202 \pm 52 $ & 205 & -21 \\ 
ComaBerenices & $ 0.41 \pm 0.02 $ & $ -1.71 \pm 0.02 $ & $ 216 \pm 4 $ &  $ -151 \pm 4 $ & 264 & 81 \\ 
Crater2 & $ -0.07 \pm 0.02 $ & $ -0.11 \pm 0.01 $ & $ 74 \pm 11 $ &  $ 71 \pm 6 $ & 102 & -80 \\ 
DESJ0225+0304 & $ 1.07^{+0.58}_{-0.56} $ & $ -0.21^{+0.48}_{-1.21} $ & $ -41^{+65}_{-63} $ &  $ 153^{+54}_{-136} $ & 159 & --- \\ 
Draco & $ 0.042 \pm 0.005 $ & $ -0.19 \pm 0.01 $ & $ 145 \pm 0 $ &  $ -56 \pm 4 $ & 156 & -88 \\ 
Draco2 & $ 1.08 \pm 0.07 $ & $ 0.91 \pm 0.08 $ & $ 268 \pm 7 $ &  $ 132 \pm 8 $ & 299 & -164 \\ 
Eridanus2 & $ 0.12 \pm 0.05 $ & $ -0.06 \pm 0.05 $ & $ 57 \pm 90 $ &  $ -4 \pm 90 $ & 57 & -73 \\ 
Eridanus3 & $ 1.08^{+0.14}_{-0.45} $ & $ -0.49^{+0.17}_{-0.13} $ & $ 284^{+58}_{-186} $ &  $ -76^{+70}_{-54} $ & 294 & --- \\ 
Fornax & $ 0.382 \pm 0.001 $ & $ -0.359 \pm 0.002 $ & $ 102 \pm 0 $ &  $ -99 \pm 0 $ & 142 & -37 \\ 
Grus1 & $ 0.07 \pm 0.05 $ & $ -0.29^{+0.06}_{-0.07} $ & $ -43 \pm 28 $ &  $ 57^{+34}_{-40} $ & 71 & -187 \\ 
Grus2 & $ 0.38 \pm 0.03 $ & $ -1.46 \pm 0.04 $ & $ 46 \pm 8 $ &  $ -131 \pm 10 $ & 139 & -132 \\ 
Hercules & $ -0.03 \pm 0.04 $ & $ -0.36 \pm 0.03 $ & $ 130 \pm 25 $ &  $ -66 \pm 19 $ & 146 & 150 \\ 
Horologium1 & $ 0.82 \pm 0.03 $ & $ -0.61 \pm 0.03 $ & $ 146 \pm 11 $ &  $ -126 \pm 11 $ & 193 & -32 \\ 
Horologium2 & $ 0.76^{+0.2}_{-0.29} $ & $ -0.41^{+0.23}_{-0.21} $ & $ 119^{+74}_{-107} $ &  $ -49^{+85}_{-78} $ & 128 & 22 \\ 
Hydra2 & $ -0.34 \pm 0.1 $ & $ -0.09^{+0.08}_{-0.09} $ & $ -88 \pm 64 $ &  $ 41^{+51}_{-57} $ & 97 & 123 \\ 
Hydrus1 & $ 3.79 \pm 0.01 $ & $ -1.5 \pm 0.01 $ & $ 333 \pm 1 $ &  $ -146 \pm 1 $ & 363 & -91 \\ 
Indus1 & $ 0.36 \pm 0.07 $ & $ -0.85 \pm 0.06 $ & $ 160 \pm 33 $ &  $ -164 \pm 28 $ & 229 & --- \\ 
Leo1 & $ -0.05 \pm 0.01 $ & $ -0.11 \pm 0.01 $ & $ -8 \pm 12 $ &  $ 75 \pm 12 $ & 75 & 169 \\ 
Leo2 & $ -0.14 \pm 0.02 $ & $ -0.12 \pm 0.02 $ & $ -61 \pm 22 $ &  $ 83 \pm 22 $ & 103 & 21 \\ 
Leo4 & $ -0.08 \pm 0.09 $ & $ -0.21 \pm 0.08 $ & $ 46 \pm 66 $ &  $ 23 \pm 58 $ & 52 & 5 \\ 
Leo5 & $ -0.06 \pm 0.09 $ & $ -0.25^{+0.09}_{-0.08} $ & $ 48 \pm 84 $ &  $ -49^{+84}_{-74} $ & 69 & 54 \\ 
LeoT & $ -0.01 \pm 0.05 $ & $ -0.11 \pm 0.05 $ & $ 9 \pm 99 $ &  $ 0 \pm 99 $ & 9 & -63 \\ 
Pegasus3 & $ 0.06 \pm 0.1 $ & $ -0.2 \pm 0.1 $ & $ -4 \pm 97 $ &  $ -31 \pm 97 $ & 32 & -56 \\ 
Phoenix & $ 0.09 \pm 0.03 $ & $ -0.07 \pm 0.03 $ & $ 17 \pm 58 $ &  $ 22 \pm 58 $ & 28 & -114 \\ 
Phoenix2 & $ 0.48 \pm 0.04 $ & $ -1.17 \pm 0.05 $ & $ 81 \pm 16 $ &  $ -258 \pm 20 $ & 271 & -41 \\ 
Pictor1 & $ 0.16 \pm 0.08 $ & $ 0.0 \pm 0.1 $ & $ -58 \pm 44 $ &  $ 56 \pm 54 $ & 80 & --- \\ 
Pictor2 & $ 1.17^{+0.1}_{-0.08} $ & $ 1.12 \pm 0.07 $ & $ 166^{+22}_{-17} $ &  $ 213 \pm 15 $ & 270 & --- \\ 
Pisces2 & $ 0.11 \pm 0.11 $ & $ -0.24^{+0.12}_{-0.11} $ & $ 11 \pm 95 $ &  $ -45^{+104}_{-95} $ & 47 & -69 \\ 
Reticulum2 & $ 2.39 \pm 0.01 $ & $ -1.36 \pm 0.02 $ & $ 182 \pm 1 $ &  $ -114 \pm 3 $ & 214 & -97 \\ 
Reticulum3 & $ 0.36 \pm 0.14 $ & $ 0.05^{+0.19}_{-0.25} $ & $ -3 \pm 61 $ &  $ 78^{+83}_{-109} $ & 78 & 101 \\ 
Sagittarius2 & $ -0.77 \pm 0.03 $ & $ -0.89 \pm 0.02 $ & $ -224 \pm 10 $ &  $ -84 \pm 7 $ & 239 & -98 \\ 
Sculptor & $ 0.099 \pm 0.002 $ & $ -0.160 \pm 0.002 $ & $ -103 \pm 0 $ &  $ 126 \pm 0 $ & 163 & 77 \\ 
Segue1 & $ -2.21 \pm 0.06 $ & $ -3.34 \pm 0.05 $ & $ -189 \pm 7 $ &  $ -149 \pm 5 $ & 240 & 109 \\ 
Segue2 & $ 1.47 \pm 0.04 $ & $ -0.31 \pm 0.04 $ & $ 80 \pm 7 $ &  $ 108 \pm 7 $ & 134 & 45 \\ 
Sextans1 & $ -0.41 \pm 0.01 $ & $ 0.04 \pm 0.01 $ & $ -112 \pm 4 $ &  $ 190 \pm 4 $ & 220 & 66 \\ 
Triangulum2 & $ 0.56 \pm 0.05 $ & $ 0.07 \pm 0.06 $ & $ -80 \pm 7 $ &  $ 137 \pm 9 $ & 159 & -253 \\ 
Tucana2 & $ 0.9 \pm 0.02 $ & $ -1.26 \pm 0.02 $ & $ 165 \pm 5 $ &  $ -130 \pm 5 $ & 210 & -207 \\ 
Tucana3 & $ -0.08 \pm 0.01 $ & $ -1.62 \pm 0.02 $ & $ -126 \pm 1 $ &  $ -5 \pm 2 $ & 126 & -198 \\ 
Tucana4 & $ 0.54 \pm 0.06 $ & $ -1.67 \pm 0.07 $ & $ 3 \pm 14 $ &  $ -197 \pm 16 $ & 197 & -86 \\ 
Tucana5 & $ -0.14^{+0.06}_{-0.05} $ & $ -1.15^{+0.08}_{-0.06} $ & $ -143^{+16}_{-13} $ &  $ -111^{+21}_{-16} $ & 181 & -140 \\ 
UrsaMajor1 & $ -0.39 \pm 0.03 $ & $ -0.63 \pm 0.03 $ & $ -109 \pm 14 $ &  $ -63 \pm 14 $ & 126 & -6 \\ 
UrsaMajor2 & $ 1.72 \pm 0.02 $ & $ -1.89 \pm 0.03 $ & $ 256 \pm 3 $ &  $ -57 \pm 4 $ & 262 & -31 \\ 
UrsaMinor & $ -0.124 \pm 0.004 $ & $ 0.078 \pm 0.004 $ & $ 120 \pm 0 $ &  $ 88 \pm 0 $ & 148 & -78 \\ 
Virgo1 & $ -0.32 \pm 0.14 $ & $ -0.62^{+0.1}_{-0.08} $ & $ -20 \pm 61 $ &  $ -89^{+43}_{-35} $ & 91 & --- \\ 
Willman1 & $ 0.21 \pm 0.06 $ & $ -1.08 \pm 0.09 $ & $ 117 \pm 11 $ &
                                                                     $
                                                                     29
                                                                     \pm
                                                                     16
                                                                     $
                                                                                                                                              & 120 & 36 \\
  \hline
Eridanus2 & $ 0.21 \pm 0.09 $ & $ -0.05 \pm 0.11 $ & $ 219 \pm 162 $ &  $ 14 \pm 198 $ & 220 & -73 \\ 
LeoT & $ 0.18^{+0.28}_{-0.29} $ & $ -0.08^{+0.17}_{-0.16} $ & $ 385^{+553}_{-573} $ &  $ 59^{+336}_{-316} $ & 389 & -63 \\ 
Phoenix & $ 0.09 \pm 0.03 $ & $ -0.07 \pm 0.04 $ & $ 17 \pm 58 $ &  $ 22 \pm 78 $ & 28 & -114 \\   
\end{tabular*}
\caption{Systemic proper motion estimates for Milky Way
  satellite galaxies, using Gaia EDR3 and the methodology of
  \cite{mcconnachie2020a}. Indus 2 was also examined but no systemic
  proper motion was able to be derived. This is an update of Table 4 in
  \cite{mcconnachie2020a}, to which the reader is referred.}
\label{preferred}
\end{center} }
\end{table*}



\bibliographystyle{aasjournal}

\end{document}